\documentclass[12pt,preprint]{aastex}

\shorttitle{Hot subdwarfs in the Galactic Bulge}
\shortauthors{G.Busso et al.}

\begin{document}

\title{Hot subdwarfs in the Galactic Bulge 
  \footnote{Based on observations collected at the European Southern
  Observatory, Chile (ESO proposal 73.D-0168A) }}

\author{G.Busso\altaffilmark{1}, S.Moehler \altaffilmark{1},
  M.Zoccali\altaffilmark{2}, U.Heber\altaffilmark{3} and S.K.Yi\altaffilmark{4}}
\altaffiltext{1}{Institut f\"{u}r Theoretische Physik und Astrophysik der
  Universitaet Kiel, 24098 Kiel, Germany; busso@astrophysik.uni-kiel.de, moehler@astrophysik.uni-kiel.de}

\altaffiltext{2}{Departamento de Astronomia y Astrofisica, Pontificia
  Universidad Catolica de Chile, Avenida Vucuna Mackenna 4860, 782-0436 Macul,
  Santiago, Chile; mzoccali@astro.puc.cl}

\altaffiltext{3}{Dr.Remeis-Sternwarte, Astronommisches Institut der
  Universitaet Erlangen-Nurnberg, Sternwartstr. 7, 96049 Bamberg, Germany; heber@sternwarte.uni-erlangen.de}

\altaffiltext{4}{Department of Physics, University of Oxford, Keble Road,
  Oxford OX1 3RH, UK; yi@astro.ox.ac.uk}

\begin{abstract}
Recent observations and theories suggest that extreme horizontal branch (EHB)
stars and their progeny should be the cause of the UV excess seen in the
spectra of many elliptical galaxies. Since the Galactic
Bulge is the closest representation of an old, metal-rich spheroid in which we
are able to study the EHB scenario in detail, we obtained spectra of bulge EHB
star candidates and we confirm their status as hot evolved stars. It is the
first time that such stars are unambiguously identified in the Galactic Bulge.
\end{abstract}

\keywords{UV excess; Galactic Bulge; Extreme Horizontal Branch stars: general}
\maketitle

\section{Introduction}
The spectra of elliptical galaxies and bulge regions of spiral galaxies in
many cases show a strong and unexpected increase in flux at wavelengths shorter
than 2500~\AA. This "UV excess" was one of the most important
discoveries of satellite based UV astronomy \citep{code69} but also a puzzle,
since it requires the existence of hot stars in these old metal-rich
systems. After a long debate most people agree that the observed UV radiation is mainly produced by very hot extreme
horizontal branch stars (burning helium in their core) and their progeny, as
Post-EarlyAGB and AGB-manqu\'e stars (O'Connell 1999; Greggio \& Renzini 1990,
1999; Dorman et al. 1995; Yi et al. 1998).
This view is supported by spectroscopic (Ferguson et al. 1991; Brown et
al. 1997, 2002) and photometric \citep{brown02} UV observations of
extragalactic systems. 
Near-UV HST observations of Brown et al.~(2000) in M32 detected for
the first time individual EHB star candidates in an elliptical galaxy. 
The best fit to these observations is achieved with evolutionary tracks for 
helium- and metal-rich populations, since in this case EHB stars have the 
longest lifetimes in the temperature range required to reproduce the UV
excess.

The closest system similar to an elliptical galaxy with respect to age and
metallicity, for which it is possible to resolve stars, is the Galactic Bulge. 
The vast majority of EHB stars known in the Milky Way, however, belongs to the
metal-poor globular clusters \citep{sabine01} or to the disk population, where
they show up as so-called subdwarf B [sdB] stars \citep{heber86,saffer94,ville95,altmann04}

The first sdB candidates in the bulge were found in the two massive
globular clusters NGC 6388 and NGC 6441 \citep{rich97,io04}
which are, however, not typical for the bulge population. The situation
changed recently: the imaging surveys of both Terndrup et al. (2004) and
Zoccali et al. (2003) of bulge fields show a sequence of hot stars that are
good candidates for EHB stars (see Fig.~1). 

\subsection{Disk Stars or Bulge Stars?}
As the line-of-sight towards the Galactic bulge passes through the disk it is 
important to estimate the expected number of disk sdB stars in the
observations. To do so we used the values of Villeneuve et al.~(1995) 
for the space density of local field sdB stars (2--4 $\times$ 10$^{-7}$
pc$^{-3}$) to derive the expected number of sdB stars along our
line of sight from 4.5 kpc (corresponding to $I\approx$ 18.5) to 11 kpc 
($I\approx$ 20.5) within the field of view of the Wide Field Imager (WFI, 30\arcmin
$\times$ 30\arcmin). The sdB stars in the field of the Milky Way consist of a
mixture of thin and thick disk stars (e.g. Altmann et al.~2004), so we
assumed a ratio of 50:50. For the thin disk we used a scale height of
325~pc and a scale length of 3.5 kpc and for the thick disk we used
values of 900 pc and 4.7 kpc, respectively \citep{larsen03}. 
We used a distance to the Galactic center of 8.5~kpc and assumed that 
the disk ends at a radial distance of 1~kpc from the Galactic Center (Robin et
al.~2003). 
This way we predict a total number of 4 to 9 sdB foreground stars within the
full field of WFI. Using the values of Ojha (2001) for the scale lengths
(2.8~kpc and 3.5~kpc) instead, we expect 7 to 14 foreground sdB stars.
However, we detect many more bulge candidates (about 140) in the WFI
photometry of Zoccali et al.~(2003).
Since they could be cool foreground stars with low reddening
(instead of reddened hot stars), we
obtained spectroscopy of 29 candidates in order to derive effective
temperatures and surface gravities and then, by means of comparison
with HB models, to check their evolutionary status. 

\section{Observations and Data Reduction}

Our spectroscopic targets were selected from the photometric catalogue of
bulge stars obtained from Zoccali et al.~2003 (see Fig.~1). The field is
located toward the Galactic center, at $l$=$0^{\circ}, b$=$-6^{\circ}$, where
the average reddening is E$_{\rm B-V}$=0.45 (Zoccali et al.~2003).   
We have chosen the stars with $18 < I < 21$ and $0 < V-I <
0.8$ and among them we selected the most isolated ones. 
After positioning as many slitlets as possible on EHB star
candidates, the remaining ones were used to get spectra of cool main
sequence and red giant stars that will be helpful for constructing the
overall spectrum expected for these bulge regions.

We obtained medium-resolution spectra ($R \approx 1200$) of 29 EHB
star candidates at the VLT-UT1 (Antu) with FORS2. We used the multi-object
spectroscopy (MXU) mode of FORS2 with the standard collimator, a slit width of
0.7" and grism B600, which allows to obtain spectra in the range between 3650
and 5200 \AA  ~(not all candidates though have full spectral coverage because
of the different positions on the CCD).  

The data reduction was performed as described in Moehler et al.~(2004)
except for the following points. Due to the long exposure times (from 2700s to
5400s) the scientific observations contained a large number of cosmic rays and
were therefore corrected with the algorithm described in Pych~(2004).
Regarding the subtraction of the sky background, we used two different
methods depending on whether the target star in the slitlet was isolated or
not. If the star was isolated, meaning any other stars in the slitlets were
well enough separated from our target to identify regions uncontaminated by
any stellar source, we approximated the spatial distribution of the sky
background by a constant. If the slitlet showed severe crowding, meaning that
the spectra of different stars were overlapping, we fitted each stellar
profile with a Lorentzian function so  that the whole spatial profile was
reproduced by the sum of all the profiles; all profiles but that one of
the target were then subtracted (for details see Moehler \& Sweigart, 2006). 
With the extraction of the spectra, we saw that some (5 of 29) of our targets
were actually cool stars. Therefore we did not proceed further with the
reduction for these stars. 
The spectra were flux calibrated using standard star spectra and corrected
for any Doppler shifts determined from Balmer lines, as in Moehler et
al.~(2004).  

\section{Analysis}

Some examples of the spectra are shown in Fig.~2. The spectra of
the hot stars show evidence for high reddening like a strong
\ion{Ca}{2} K line and the diffuse interstellar band at 4430~\AA.
To fit the spectra (except for one He-rich star) we used ATLAS9 model atmospheres
for solar metallicity (Kurucz 1993) to account for effects of radiative
levitation (see Moehler et al.~2000 for details), from which we calculated
spectra with Lemke's version of the
LINFOR program (developed originally by Holweger, Steffen, and Steenbock at
Kiel University). The use of NLTE models or of LTE models with higher 
metallicity does not significantly change the results.

To establish the best fit, we used the routines developed by Bergeron
et al.~(1992) and Saffer et al.~(1994), as modified by Napiwotzki et
al.~(1999), which employ a $\chi^2$ test. The uncertainty necessary for the
calculation of  $\chi^2$ is estimated from the noise in the continuum
regions of the spectra. The fit program normalizes model spectra and
observed spectra using the same points for the continuum definition.
We used the Balmer lines $H_{\beta}$ to $H_{10}$ (excluding
$H_\epsilon$ to avoid the Ca~II H line), the He~I lines  at  4026,
4388, 4471, 4921 \AA,  and the He~II lines at 4542 and 4686 \AA.  

We obtained the atmospheric parameters $T_{eff}$, log$g$ and helium
abundances and we calculated the absolute V and I magnitudes expected
for these values, assuming $M_{star}=0.5M_{\odot}$. We left out one
star because the fit was unacceptably poor.  Since the formal fit
errors are underestimated by a factor 2--4 (Napiwotzki, priv. comm.) a
formal error of 0.1 in log$g$ implies an error of 25\%--50\% in the
distance. We therefore discuss the bulge membership only for those
stars with a formal error in log$g$ of less than 0.1 because for the
others (4 of 23) the uncertainty in the distance is too large. Then
considering a distance from the Galactic Center of $\approx$ 8.5~kpc
and a bulge radius of $\approx$ 1.5 kpc, we find that most of these
objects are indeed bulge stars: of 19 hot stars with reasonable errors $\sigma$
in the distance, 13 stars are in the bulge within 1 $\sigma$ and 3
more are in the bulge within 3 $\sigma$. We thus found 3 probable disk EHB 
stars in our sample of 29 candidate EHB stars, which corresponds to
a contamination of 10\%.  

The heliocentric radial velocities also suggest a bulge membership of
most EHB stars. The field where we are looking is at Galactic
coordinates $l$=$0^{\circ}, b$=$-6^{\circ}$, toward the Galactic
center, i.e. the expected radial velocities for disk stars are around
0 km~s$^{-1}$. Our velocities are distributed in a range between
$-$200 and $+$300~km~s$^{-1}$, in agreement with the values found for
K giants in Baade's Window by Terndrup et al. (1995, between $-$240
and $+$194~km~s$^{-1}$). We calculated a velocity dispersion of 110
$\pm$ 17 km s$^{-1}$ from our bulge EHB stars: the expected value for
the disk is 50--70 km s$^{-1}$ (Lewis \& Freedman 1989) while Terndrup
et al. (1995) found for Baade's Window, at a distance of 8kpc, 80--110
km s$^{-1}$.

Finally we compare our results with horizontal branch theoretical tracks: in
Fig.~3 we plot the values found for those stars which belong to the bulge in
the ($T_{eff}$, $\log g$) diagram. The error bars are the
formal errors from the fit procedure, but, as we already mentioned, these
errors are underestimated and in addition, they do
not include any systematic uncertainties, due to, e.g. sky
subtraction, flux calibration, etc.
The evolutionary tracks are from Yi et al.~(1997) with metallicity Z=0.004 and
helium abundance Y=0.2416. 
The Zero Age Horizontal Branch (ZAHB), where the star starts to burn helium in
its core quietly, and the Terminal Age Horizontal Branch (TAHB), where the star
has burned the 99\% of the helium, are shown together with evolutionary tracks 
for stars with total masses of 0.49, 0.50, and 0.51 $M_{\odot}$ (respectively
$M_{env}=$ 0.0075, 0.0127, 0.0226 $M_{\odot}$). 
Our tracks end at 0.495 $M_{\odot}$, corresponding to $M_{env}=$ 0.0075 $M_{\odot}$; since EHB stars may have $M_{env} <$ 0.005
$M_{\odot}$, we extrapolated the ZAHB and TAHB to higher
temperatures (dashed curves in Fig.~3) to guide the eye.
Proper models for lower envelope masses will be calculated and used in
a later paper. 
The observed points agree quite well with the theoretical tracks, therefore
these objects are indeed EHB stars; some objects are above the TAHB
meaning that they are in the post-HB phase and then evolving as AGB-manqu\'e
stars (Greggio \& Renzini~1990). 

Finally we want to mention that all stars except one (which is helium-rich)
are helium deficient as expected from diffusion. 

\section{Conclusions}

We observed spectra of 29 EHB star candidates in the Galactic Bulge, from which
we estimated the  atmospherical parameters $T_{eff}$ and log$g$ to verify
their evolutionary status. Five objects turned out
to be cooler foreground stars with low reddening, and for another one
the spectroscopic fit is unacceptably bad. Of the 19 hot
stars with reasonable distance errors
16 lie within a radius of 1.5~kpc around the Galactic center at
8.5~kpc. Also the observed radial velocities support a membership of
these stars to the bulge.
This is the first time that of stars are observed in the bulge and 
our statistics show that either spectroscopy or multi-colour photometry is 
required to disentangle hot stars in the bulge from other sources. 
We will use these spectra and the parameters derived from them
to construct the integrated spectrum of the galactic bulge from the UV to the
optical, following the method of Santos et al.~(1995).
This study will verify the role, so far only predicted, of these stars
regarding the UV excess in the elliptical galaxies.

\acknowledgments
We are grateful to the ESO staff, especially those at the Paranal
observatory, for all their help with the observations. We thank an
anonymous referee for pertinent comments, which helped us to improve
this paper. GB acknowledges support from the Deutsche
Forschungsgemeinschaft via grant Mo 602/8.

\clearpage

\begin{figure}
\epsscale{.90}
\plotone{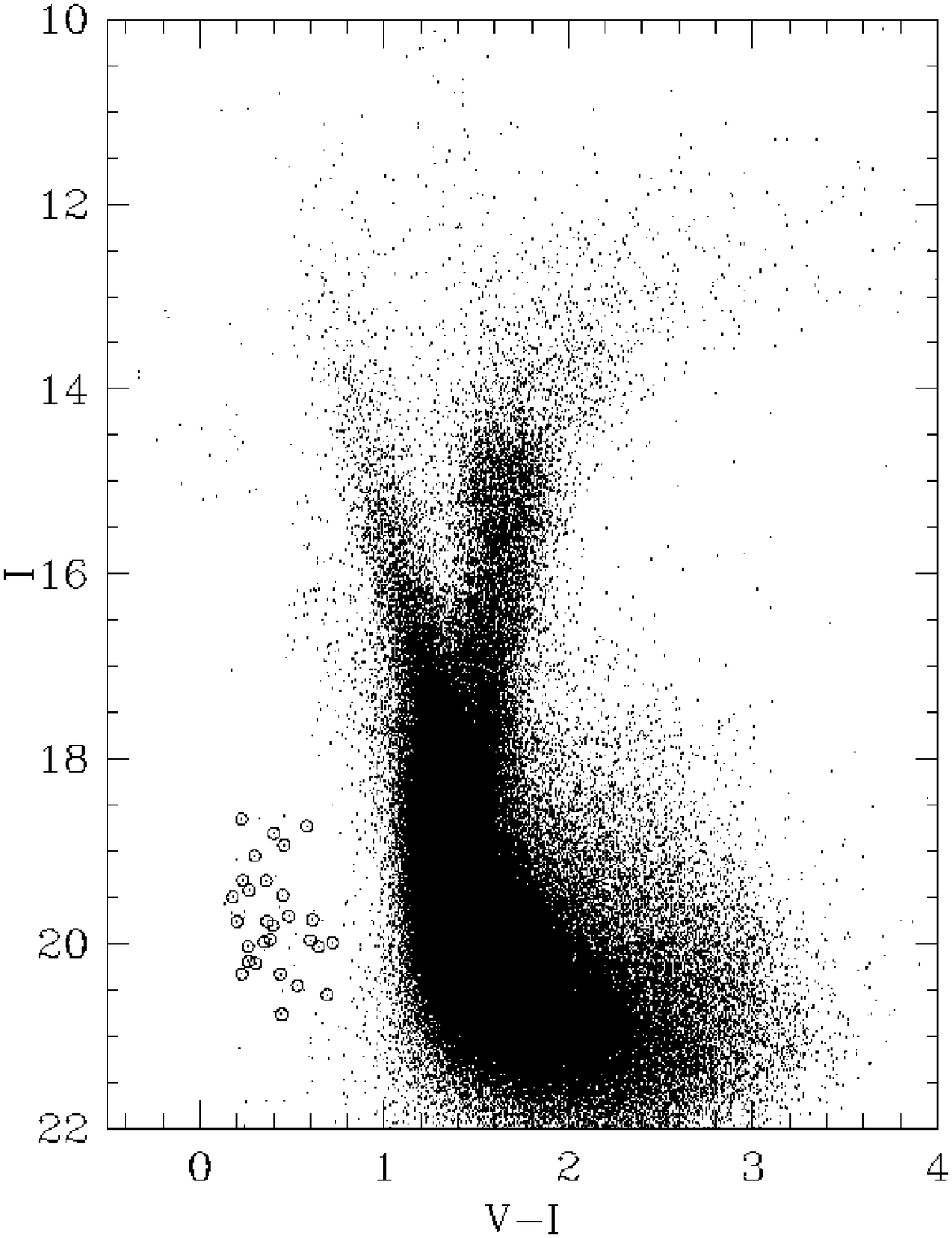}
\caption{Colour-magnitude diagram of the Galactic bulge (at $l$=$0^{\circ},
  b$=$-6^{\circ}$,  E$_{\rm B-V}$=0.45) obtained from the
  Zoccali et al. (2003) observations. Our targets are marked with circles.}
\end{figure}

\clearpage

\begin{figure}[htbp]
\epsscale{1.100}
\plotone{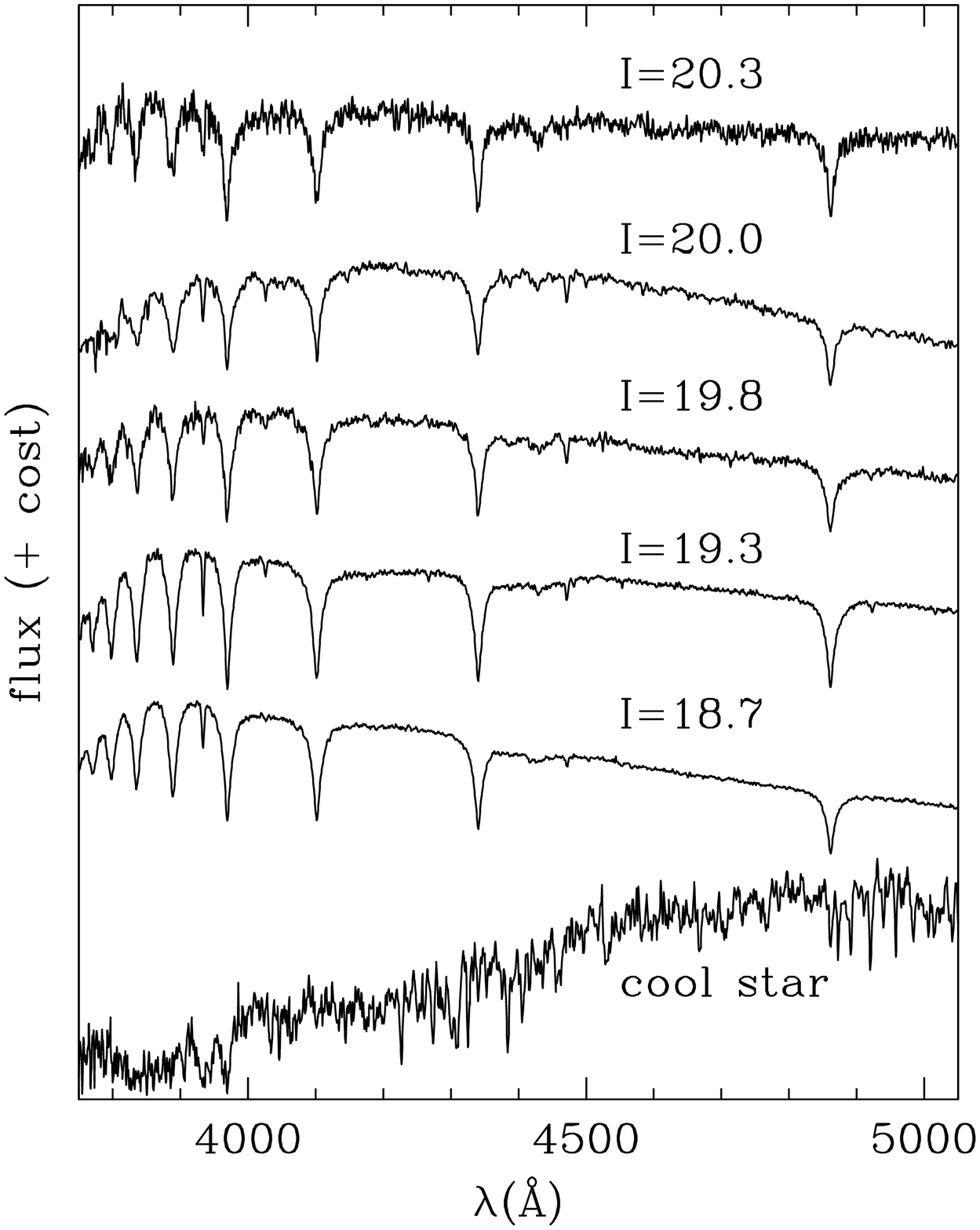}
\caption{Examples of the spectra of the sdB star candidates: the spectrum at
the bottom is for a candidate which turned out to be a cool star and shown here for
comparison; the others are typical spectra we found, indicating hot stars with strong Balmer lines.} 
\end{figure} 

\clearpage

\begin{figure*}[htbp]
\begin{center}
\includegraphics[angle=270,scale=.40]{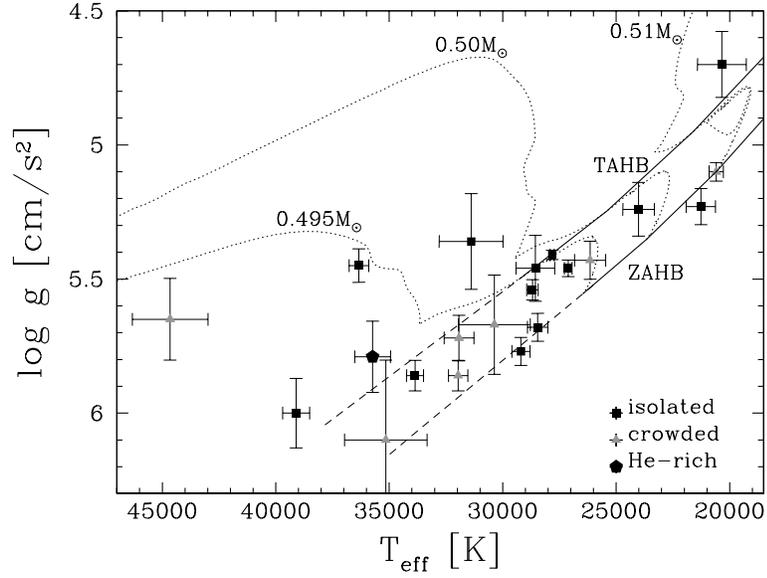}
\end{center}
\caption{($T_{eff}$, $\log g$) diagram: the squares indicate the
  isolated stars; the triangles indicate the crowded stars and the
  pentagon is the He-rich star. The ZAHB and TAHB \citep{yi97} for
Z=0.004 and Y=-0.2416 are plotted together with
  evolutionary tracks for 0.495, 0.50 and 0.51 $M_{\odot}$. The dashed lines are
  extrapolated from the ZAHB and TAHB tracks.}  
\end{figure*}

\end{document}